\documentclass[aps,amsmath,graphicx,epsf,rotate]{revtex4}
\usepackage{graphicx}
\usepackage{comment}
\usepackage{amsthm}
\usepackage{amsfonts}
\gdef\ffrac#1#2{\textstyle{\frac{#1}{#2}}\displaystyle}
\gdef\be{\begin{equation}}
\gdef\ee{\end{equation}}

\gdef\p{{\partial}}
\begin{document}
%
\title[Yang-Lee Edge]{The Yang-Lee Edge Singularity and Related Problems
\footnote{Contribution to {\sl Fifty years of the renormalization group. Dedicated to the memory of Michael E.~Fisher},
 A.~Aharony, O.~Entin-Wohlman, D.~Huse and L.~Radzihovsky, eds. (World Scientific, to appear.)}}
\author{John Cardy}
\affiliation{Department of Physics, University of California, Berkeley CA 94720, USA\\
and All Souls College, Oxford OX1 4AL, UK.}
\email{john.cardy@all-souls.ox.ac.uk}

\begin{abstract}
The Yang-Lee edge singularity is a prototypical example of the application of renormalization group ideas to critical behavior, and one to which Michael Fisher made several important contributions. Moreover it has connections to several other problems such as the statistics of branched polymers, and its scaling limit in two dimensions provides a simple example of integrable field theory. This article aims to give a pedagogical introduction to these matters, with a few new ideas thrown in. 
\end{abstract}
\maketitle

\section{Introduction}\label{secintro}
In 1952 T.D.~Lee and C.N.~Yang published two remarkable back-to-back papers in \em The Physical Review \em\cite{YL,LY}.\footnote{Since the author order is reversed between the two papers, in the literature this work is referred to interchangeably as Yang-Lee theory or Lee-Yang theory. In this chapter I shall follow Fisher and use the first option.}
Exploiting the equivalence between the ferromagnetic Ising model in an external magnetic field $h$ and a lattice gas with activity $z=e^{h/kT}$, they proved a theorem which states that, on any finite graph of degree $N$, the partition function $Z$, which is proportional to a polynomial in $z$ of degree $N$, has its zeroes only on the unit circle $|z|=1$, corresponding to purely imaginary $h$. Moreover since $Z$ is a sum of positive terms for real $h$, these zeroes are bounded away from the real axis. They argued that this should persist at sufficiently high temperatures in the thermodynamic limit $N\to\infty$ on suitably regular lattices with finite-range interactions, and that the zeroes should become dense along the imaginary axis for $|{\rm Im}\,h|$ greater than some critical value $h_c$. This limit point  is known as the Yang-Lee edge. They argued that mechanism for the free energy per site
$\lim_{N\to\infty}(-N^{-1})\log Z$ in zero field becoming singular as $T$ approaches  the critical temperature $T_c$ from above is that the Yang-Lee edge approaches the real axis, that is $h_c\to0$.

At the time this must have seemed a promising idea. Very little was understood theoretically about critical behavior beyond Curie-Weiss theory, Onsager's result for the free energy of the two-dimensional Ising model in zero field\cite{onsager} and Yang's expression for the spontaneous magnetization.\cite{yangmag} However, although locating Yang-Lee zeros remains a useful numerical tool for many varied problems, its analytic power has turned out to be less impactful than the understanding gained from the spectrum of the transfer matrix in integrable models. Compared with renormalization group ideas, it gives little insight into the origin of anomalous scaling behavior.

Almost twenty years later Kortman and Griffiths \cite{kg} argued, on the basis of series expansions, that the density of zeroes has a power law behavior near the edge, namely $\rho(h')\propto(h'-h_c)^\sigma$ as the imaginary part $h'\downarrow h_c$. They estimated $\sigma\approx-0.1$ for the square lattice and $\approx+0.1$ for the three-dimensional diamond lattice. Yang and Lee had already showed that $\sigma=-\ffrac12$ for a one-dimensional model (see Sec.~\ref{sec1d}), while $\sigma=\ffrac12$ in mean field theory. 

However it was Fisher \cite{mefyl} who recognized that the Yang-Lee edge singularity is a \em bona fide \em example of critical behavior, with a diverging correlation length and universal critical exponents 
which depend only on the dimensionality $d$ of space. The only feature that makes it unusual is that, because of the imaginary magnetic field, the Boltzmann weights are no longer real and non-negative, so do not give a probability measure. This makes some properties, for example, the arguments for the existence of the thermodynamic limit, questionable. However this has not stopped theoretical physicists from embracing it as one of the simplest examples of a critical point with a diverging correlation length, and unleashing the arsenal of modern theoretical methods upon it. It is the aim of this Chapter to describe some of these. It turns out that the problem is an ideal classroom for understanding of some of the fundamental ideas of the last fifty years of theoretical statistical physics, many of which bear Michael Fisher's mark.

\section{The Yang-Lee Edge}\label{secyl}
\subsection{Yang-Lee Theorem}
In this section we give a brief account of the Yang-Lee theorem and its proof, since the original papers are not easy to follow for a modern reader, and some of the accounts given in the literature are overly complicated or incomplete. In fact there are at least two different lines of approach, algebraic or analytic, with different generalizations. We give a simplified version of the latter, due to Ruelle \cite{rue}.

Suppose we have a finite graph $\cal G$ with vertices labelled by $i=1,\ldots,N$ and edges by $(jk)$ with $j<k$. At each vertex $i$ there is an Ising spin $s_i$ taking the values $\pm1$. The partition function is
\be
Z^{\cal G}(h_1,\ldots,h_N)=\sum_{\{s_j\}}e^{\beta\sum_{jk}V_{jk}s_js_k+\beta\sum_jh_js_j}\,,
\ee
where we have allowed for a different external magnetic field $h_j$ at each vertex. This may be mapped to a lattice gas by defining occupation numbers $n_j=\frac12(s_j+1)$ taking the values $0$ or $1$. Restricting $n\leq1$ then models an on-site repulsion, while taking $V_{jk}\geq0$ corresponds to a general attractive two-body potential (not necessarily nearest neighbor) for the lattice gas, or a ferromagnetic interaction in the Ising model. In the gas picture we may think of $Z$ as being proportional to the grand partition function
\be\label{Xi1}
\Xi^{\cal G}(z_1,\ldots,z_N)=\sum_{\{n_j=0,1\}}\prod_jz_j^{n_j}e^{-4\beta\sum_{jk}V_{jk}(n_j-n_k)^2}\,.
\ee
where $z_j=e^{2\beta h_j}$ is now interpreted as the local activity of the gas.
We have written this in a form where it is a multinomial of degree one in each of its arguments $z_j$. Under the Ising symmetry $(s_j\to-s_j, h_j\to-h_j)$, which translates into $(n_j\to1-n_j, z_j\to z_j^{-1})$, 
\be\label{inv}
\Xi^{\cal G}(z_1,\ldots,z_N)=(\prod_{j=1}^Nz_j)\,\Xi^{\cal G}(1/z_1,\ldots,1/z_N)\,.
\ee
Note also that, as long as $V_{jk}$ is real, 
\be\label{real}
\Xi^{\cal G}(\{z_j\})=\overline{\Xi^{\cal G}}(\{\bar z_j\})\,.
\ee

The first part of the proof is to show that

\noindent{\bf Proposition.}
\em $\Xi^{\cal G}(z_1,\ldots,z_N)$ is non-zero in $\cap_{j=1}^N\{|z_j|<1\}$.\em
\noindent The proof proceeds by induction on $N$. Suppose it is true for all graphs of degree $\leq N-1$. Considering the sum over $n_1$ in (\ref{Xi1}), the term with $n_1=0$ just gives the partition function on ${\cal G}\setminus\{1\}$, while that with $n_1=1$ modifies the activities at the remaining vertices of ${\cal G}\setminus\{1\}$. Explicitly, setting $a_{jk}\equiv e^{-\beta V_{jk}}\leq1$,\footnote{Although if $V_{jk}$ is a nearest neighbor interaction, the repulsive case $V_{jk}<0$ on a bipartite lattice may be mapped to the ferromagnetic case by reversing the sign of  $s_j$ on one sublattice, this is special and the Yang-Lee theorem does not generally apply to repulsive interactions.}
\be
\Xi^{\cal G}(z_1,z_2,z_3,\ldots)=\Xi^{{\cal G}\setminus\{1\}}(a_{12}z_2,a_{13}z_3,\ldots)+
z_1a_{12}\ldots a_{1n}\Xi^{{\cal G}\setminus\{1\}}(z_2/a_{12},z_3/a_{13},\ldots)
\ee
The first term is non-zero by hypothesis, and, using (\ref{inv}) and (\ref{real}),  the second may be written
\be
z_1z_2\ldots z_n
\overline{\Xi^{{\cal G}\setminus\{1\}}}(a_{12}/\bar z_2,a_{13}/\bar z_3,\ldots)
\ee
so we need to show that this smaller in modulus than the first. A sufficient condition is
\be
\sup_{|a_{1j}|\leq1}\sup_{|z_j|<1}\left|\frac{z_1z_2\ldots z_n
\Xi^{{\cal G}\setminus\{1\}}(a_{12}/\bar z_2,a_{13}/\bar z_3,\ldots)}{\Xi^{{\cal G}\setminus\{1\}}(a_{12}z_2,a_{13}z_3,\ldots)}\right|\leq1\,.
\ee
By continuity, the left-hand side does not change if we strengthen the conditions $|a_{1j}|\leq1$
to the open set $|a_{1j}|<1$. Then, by the maximum modulus principle, which states that the supremum is attained on the boundary, the left-hand side does not change
if we replace the condition $|z_j|<1$ by $|z_j|=1$.
 However, with these modified conditions the
expression to be maximized is a constant 1, and the bound is satisfied trivially. The case $N=1$ is trivial, and, for $N=2$, $\Xi^{\cal G}(z_1,z_2)=1+a_{12}(z_1+z_2)+z_1z_2$, which vanishes when $z_2=-(1+a_{12}z_1)/(a_{12}+z_1)$. This maps $|z_1|<1$ into $|z_2|>1$, so the proposition holds.
This completes the inductive proof.

We note in passing that the inhomogeneous case has application to the problem of a lattice gas in a random medium, equivalent to the Ising model in a random field. However for the present we specialize to the case when all the $z_j$ are equal to $z$, and $\Xi^{\cal G}$ is a polynomial of degree $N$ in $z$.
We then have

\noindent {\bf Theorem [Yang-Lee].} \em The zeroes of the $\Xi^{\cal G}(z)$ all lie on the unit circle $|z|=1$.\em 
\noindent From the Proposition the interior of the unit circle is free of zeroes. But from (\ref{inv}) the exterior of the unit circle is also zero-free. This completes the proof.

 Since $\Xi(z)$ is a polynomial of degree $N$, it is determined up to a constant by its zeroes
 $\{e^{ i\theta_j}\}$. If, for example, we are interested in the mean density (magnetization),
 \be
 N\langle n\rangle=z\partial_z\log\Xi=z\partial_z\sum_{j=1}^N\log(1-e^{-i\theta_j}z)
 =\sum_{j=1}^N\frac z{z-e^{i\theta_j}}\,.
\ee

It is remarkable the density of Yang-Lee zeroes contains in principle all the information about the equation of state of the Ising model in a magnetic field.
As Fisher was fond of pointing out, all the existing approximations to the equation of state failed to take the Yang-Lee singularity into account. More recently, however Xu and Zamolodchikov\cite{xz} have made progress in this direction. 

\subsection{One dimension}\label{sec1d}
A nice feature of the Yang-Lee singularity is that in one dimension it has non-trivial behavior, and is easily solvable as an exercise. Suppose we have a lattice gas with nearest neighbor attraction, so that
\be
\Xi=\sum_{\{n_i=0,1\}}\prod_{j=0}^{L-1}z^{n_j}e^{-v(n_-n_{j+1})^2}\,,
\ee
where, for convenience, we may impose periodic boundary conditions $n_L=n_0$. This may then be written as ${\rm Tr}\,{\cal T}^L$ where $\cal T$ is the transfer matrix
\be
{\cal T}=\begin{pmatrix}1&w\\ zw &z \end{pmatrix}\,,
\ee
where $w=e^{-v}<1$, with eigenvalues
\be
\lambda_\pm=\ffrac12[1+z\pm\sqrt{(1-z)^2+4w^2z}]\,,
\ee
so that $\Xi(z)=\lambda_+^L+\lambda_-^L$. The zeroes are where
$\lambda_+/\lambda_-=e^{(2j+1)\pi i/L}$ with $j=0,\ldots,L-1$. After some algebra one finds their location to be $z=e^{i\theta_j}$, where
\be
\cos\theta_j=(1-w^2)\cos[(2j+1)\pi/L]-w^2\,.
\ee
We see that at high temperature ($w\to1$) the zeroes cluster around $\theta=\pi$, while at low temperatures ($w\to0$) they are more or less uniformly spaced around the unit circle. 

The Yang-Lee edge singularity is where the correlation length $\xi\sim\log(\lambda_+/\lambda_-)$ diverges, that is $\cos\theta_c=1-2w^2$. Near the edge 
\be
\theta_j-\theta_c\propto[(2j+1)\pi/L]^2\,,
\ee
so that in the thermodynamic limit the density of zeroes diverges as $(\theta-\theta_c)^{-1/2}$.

For  general complex $w$, the locus of Yang-Lee singularities is 
\be
z=1-2w^2\pm2w(w^2-1)^{1/2}\,.
\ee
In the repulsive case $w>1$, therefore, there are singular points on the negative real $z$ axis, the one closer  to the origin moving from $z=-1+$ to $z=0-$ as $w$ increases from $1+$ to $\infty$. This is an example of the \em repulsive gas singularity\em. Since, in this case it was found by analytic continuation from the Yang-Lee edge singularity, has the same critical behavior. This, as we shall see, should hold more generally. 

Although above we considered the simple case of the spin $\frac12$ Ising model, in one dimension the transfer matrix method may be applied to higher spin and also to models with continuous symmetries like O$(n)$. Fisher\cite{mefyl1d} gave convincing arguments that the Yang-Lee singularity is always the result of the meeting and annihilation of the two largest eigenvalues, and therefore has the same critical exponents.

\section{Renormalization Group Analysis}\label{secrg}
We now come to Fisher's most celebrated contribution\cite{mefyl} to this subject: the realization that the scaling behavior of the Yang-Lee edge singularity is given by a scalar field theory with a $\phi^3$ interaction and a purely imaginary coupling, and the application of the renormalization group ideas, including the $\epsilon$-expansion, that he and Wilson had developed for the Ising critical point ($\phi^4$  theory) six years previously.\cite{wf} I shall describe his approach, then set this in the context of renormalized field theory.
\subsection{Field theory formulation}
The most systematic way of arriving at a field theory from a lattice Ising model is through a Hubbard-Stratonovich transformation, which is simply a gaussian integral. Starting from the partition function on a regular lattice, whose vertices are labeled by a $d$-dimensional vector $r$, we write
\begin{eqnarray}
Z&=&{\rm Tr}_{\{s(r)=\pm1\}}e^{\frac12\sum_{r,r'}s(r)V(r-r')s(r')+h\sum_rs(r)}\\
&\propto&{\rm Tr}_{\{s(r)=\pm1\}}\int[d\phi(r)]e^{-\frac12\sum_{r,r'}\phi(r)V^{-1}(r-r')\phi(r')+\sum_r(h+\phi(r))s(r)}\\
&=&\int[d\phi(r)]e^{-\frac12\sum_{r,r'}\phi(r)V^{-1}(r-r')\phi(r')+\sum_r\log\cosh(h+\phi(r))}\,,\label{hs}
\end{eqnarray}
where $V^{-1}$ is the inverse of the matrix with rows and columns labeled by $r$ and $r'$ and elements $V(r-r')$. Shifting $\phi\to\phi-h$ removes the $h$-dependence in the last term and generates a linear coupling to $\phi$.    If $h$ is purely imaginary, so is this coupling. 

The inverse is easily taken in terms of the Fourier transforms $V_q$ and $\phi_q$. So far, this is exact, but now we start expanding in powers of $q$ and the fluctuations about the minimum of the potential in  $\phi$. Denoting the shifted field by $\psi$, this leads to terms of $O(\psi^2,\psi^3,\ldots)$. In real space the resulting effective free energy functional may be written as
\be\label{phi3}
{\cal H}=\int[\ffrac12((\nabla\psi)^2+r\psi^2)-iw\psi^3]d^dr\,.
\ee
Although the details of the dependence of $r$ and $w$ on the original couplings are tedious, they are analytic and only the general form of (\ref{phi3}) is important. In Fourier space\cite{mefyl}
\be
{\cal H}=h\hat\psi_0-\ffrac12\int_q(r+q^2)\hat\psi_q\hat\psi_{-q}-iw\int_q\int_{q'}
\hat\psi_q\hat\psi_{q'}\hat\psi_{-q-q'}\,.
\ee
In principle the $q$ integrals run over the first Brillouin zone. However, anticipating that rotational invariance emerges in the scaling limit near the critical point, this is replaced by $|q|<\Lambda=O(a^{-1})$ where $a$ is the lattice spacing. The neglect of all the terms discarded in reaching this point is justified \em a posteriori \em by their irrelevance (coefficients flowing to zero) under the RG to be described. 

\subsection{Momentum shell RG}\label{secmomrg}

Wilson's momentum-shell RG\cite{wilson1, wilson2} then involves (i) integrating out the wavelengths $b^{-1}\Lambda<|q|<\Lambda$ where $b>1$; (ii) rescaling $q'= bq$ so the cut-off is again $|q'|<\Lambda$. Consider first the rescaling. In order to keep the coefficient of $q^2=1$ we need to rescale $\psi_q=b^{d/2+1}\psi'_{q'}$, so then
$h'=b^{d/2+1}h$, $r'=b^2r$ and $w'=b^{\epsilon/2}w$ where $\epsilon=6-d$.

This gives the upper critical dimension for this theory, above which the RG flows go into the gaussian fixed point with $w=0$, as $d_u=6$, as compared with 4 for the $\phi^4$ theory. Note that the $\psi^3$ theory has a peculiarity which is absent from, or rather unimportant for, the $\phi^4$ case, in that, as long as $w\not=0$, one may shift the  field $\psi_{q=0}$ either to set the $r'=0$ or $h'=0$.
This is an example of the presence of a \em redundant \em field, given by the variation of the free energy with respect to $\psi_{q=0}$. Such redundant fields do not usually affect the calculation, but here one is forced to make a choice. It turns out that maintaining  the coefficient $r$ or the less relevant field gives less singular, if mathematically equivalent, RG flows. 

More importantly, it implies that the RG eigenvalues of $h$ and $r$ are not independent, as is the case with
the Wilson-Fisher fixed point: the Yang-Lee edge singularity has only one independent critical exponent (apart from corrections to scaling from irrelevant fields), all others determined by scaling relations. If the RG eigenvalue of $h$ is $y_h$, then standard scaling arguments \cite{jcbook} show that the correlation length $\xi$ diverges as $h\to h_c$ like $|h-h_c|^{-\nu}$ where $\nu=1/y_h$; that the scaling dimension\footnote{Modern approaches using conformal field theory regard the scaling dimensions $x$ of scaling fields as the fundamental data, all other critical exponents being derived from these by scaling relations.} of $\psi$ is $x_h=d-y_h$, so the two-point function at the critical point decays as $|r|^{-(d-2+\eta)}$;where $d-2+\eta=2x_h$; and the magnetization $M\sim\langle\psi\rangle\sim \xi^{-(d-y_h)}\sim |h-h_c|^{d-y_h)/y_h}$. In particular this gives the density of Yang-Lee zeroes behaving as $(h-h_c)^\sigma$ where $\sigma=(d-y_h)/y_h
=(d-2+\eta)/(d+2-\eta)$. This is Fisher's scaling relation.\cite{mefyl}

To return to Fisher's actual calculation, the next step is the elimination of the degrees of freedom with $b^{-1}\Lambda<|q|<\Lambda$.     This is most easily carried out perturbatively in $w$, and in the end it involves the same one-loop Feynman diagrams we shall encounter below in the renormalized field-theoretic  approach, so we do not give details here. It was, however, Fisher's crucial observations that (a) these diagrams depend meromorphically on $d$, and (b) the emergence, for the $\phi^4$ case, of a weak-coupling fixed point at a value of the coupling $O(4-d)$. These were the crucial ingredient in he and Wilson's discovery of the $\epsilon$-expansion as the first controlled approach to a RG calculation for a non-trivial fixed point.\cite{wf} In the 1978 paper\cite{mefyl} it must have been relatively straightforward for him to apply these methods to the $\psi^3$ theory, although by then  other authors had considered field theories with cubic interactions. 

\subsection{RG in renormalized field theory.}
The renormalized field theory approach gives a systematic way of deriving the $\epsilon$-expansion, which is perhaps less intuitive than the momentum shell approach, but  is much more efficient at higher orders, especially when combined with minimal subtraction, to be described below. 

\begin{figure}
\centering
\includegraphics[width=0.55\textwidth]{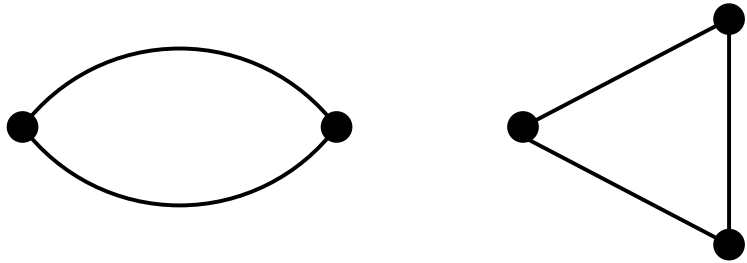}
\caption{\label{figivp}1 loop primitively divergent graphs for $\Gamma^{(2)}$ and 
$\Gamma^{(3)}$ in $\psi^3$ theory.}
\end{figure}

It is worth separating the principle of the argument from the details of the calculation. In general, an interacting statistical field theory, formally defined by  a functional integral $\int[d\psi]e^{-{\cal H}[\psi]}$, is in fact ill-defined even in perturbation theory due to ultraviolet divergences. The process of renormalization makes sense of the theory. We regard the integration field $\psi$ and the parameters $(r,w)$ in (\ref{phi3}) as 'bare' quantities $(\psi_0,r_0,w_0)$, The first step is to regulate the divergences by, for example, imposing a momentum cut off $|q|<\Lambda$, or by dimensional regularization, that is continuing in $d$ in such a way that all Feynman integrals converge, at least apart from special kinematic points. The next step is to identify by dimensional analysis which parts of which diagrams depend on the regulator. These are called the primitively divergent irreducible vertex parts. For the $\psi^3$ theory these are shown in Fig.~\ref{figivp}.
The \em statement of renormalizability\em, at least for this $\psi^3$ theory, says that \em if\em, for $d\leq6$, we do
\begin{itemize}
\item shift the bare field so that at each order $\langle\psi_R\rangle=0$;
\item mass renormalization: tune $r_0$ as a function of $w_0$ so that the renormalized mass vanishes, that is $\Gamma_R^{(2)}(q)=\langle\psi^R_q\psi^R_{-q}\rangle^{-1}$ vanishes at $q=0$ . (These first two conditions serve to put the theory on its critical manifold, and remove the worst, power-law, divergences.)
\item field renormalization: $\phi_0=Z_\psi^{1/2}\psi_R$ such that $\partial_{q^2}\Gamma_R^{(2)}(q)=1$, evaluated at some fixed but arbitrary normalization point $q^2=\mu^2$. This cannot be at $q^2=0$ because of infrared divergences for $d\leq6$; 
\item coupling constant renormalization $w_R=Z_ww_0=i\Gamma^{(3)}_R(q,q',-q-q')$ evaluated once again at an infrared finite normalization point dependent on $\mu^2$;
\end{itemize}
\em then \em, all the correlation functions
\be
G^{(N)}_R=\langle\psi_R(r_1)\ldots\psi_R(r_N)\rangle
=Z_\psi\langle\psi_0(r_1)\ldots\psi_0(r_N)\rangle\,,
\ee
when considered at a function of $w_R$, have a finite limit (at non-coincident points) as the regulator is removed, that is $\Lambda\to\infty$ or $d\to6$. In the latter case the normalization constants $Z_\psi$ and $Z_w$ have poles at $d=6$ (as well as other values of $d$ which are not important). In minimal subtraction we retain only their divergent pieces, since the rest corresponds to some finite redefinition of the renormalized theory.

The above works only if $d\leq6$: for $d<6$ there are primitive divergences in only a finite number of subdiagrams, and the theory is \em super-renormalizable\em. However there is then a proliferation of infrared divergences in the massless theory, that is, at the critical point. The $\epsilon$-expansion effectively resums these into something sensible. At  $d=6$ there are an infinite number of primitively divergent subgraphs but they occur only in $\Gamma^{(2)}$ and $\Gamma^{(3)}$, where they are logarithmic. The theory is \em exactly \em renormalizable. For $d>6$, divergences proliferate and the theory is \em non-renormalizable\em. However, the infrared behavior becomes trivial, that of the gaussian fixed point, so $d=6$ is called the upper critical dimension from the point of view of statistical physics.

Historically, in particle physics, the property of renormalizability was taken to be essential: the renormalized correlation functions are the physical ones. More recently, this point of view has altered: all field theories may be thought of as \em effective\em, valid over some large but finite range of scales. However, since we want the predictions of the theory to be insensitive to the particular value of the regulator, this is more or less equivalent to demanding renormalizability.

In statistical physics, however, the physical quantities are the bare, cut-off correlation functions. Why then does renormalizability matter? From the insensitivity of the renormalized theory to the cut-off we can say that
\be\label{ddL}
\left.\Lambda\frac{\p}{\p\Lambda}\right|_{w_R}G^{(N)}_R=\left.\Lambda\frac{\p}{\p\Lambda}\right|_{w_R}\left(Z_\psi^{-N/2}\,G^{(N)}_0\right)=0\,.
\ee
Defining the bare dimensionless coupling $g_0=w_0\Lambda^{(6-d)/2}$ so that $G^{(N)}_0$ depends on $g_0$ and $\Lambda$, we can expand out (\ref{ddL}) to get
\be
\left[\Lambda\frac{\p}{\p\Lambda}+\beta(g_0)\frac\p{\p g_0}-\frac N2\gamma(g_0)\right]G^{(N)}_0(\{q_i\})=0\,,
\ee
where $\beta(g_0)=\Lambda(\p/\p\Lambda)|_{w_R}g_0$ and
$\gamma(g_0)=\Lambda(\p/\p\Lambda)|_{w_R}\log Z_\psi$. The first term acts on the explicit $\Lambda$ dependence of $G^{(N)}_0$, the other terms on its implicit dependence on keeping $w_R$ fixed. 

These are examples of RG flow equations. In the language of Wilson's RG, if we send $\Lambda\to b^{-1}\Lambda$, with $b=1+\delta\ell$, then to keep the same long distance physics, we have $dg_0/d\ell=-\beta(g_0)$.

It remains to compute $\beta(g_0)$ and $\gamma(g_0)$, at least to one loop order.
The first diagram in Fig.~\ref{figivp} gives
\be
\Gamma^{(2)}_0(p)=p^2+\frac{(6!g_0)^2}{2}\int^\Lambda\frac{d^6q/(2\pi)^6}
{q^2(q-p)^2}\,,
\ee
where the denominator factor may be written, using Feynman parameters,
$\int_0^1[xq^2+(1-x)(q-p)^2]^{-2}dx$, which, on shifting the $q$ integration, becomes $\int_0^1[q^2+x(1-x)p^2]^{-2}dx$. In this form it is straightforward to take the derivative with respect to $p^2$ to get an expression $\propto\log\Lambda$, so that 
$\gamma(g_0)=-6K_6g_0^2+\cdots$, where $K_6$ is the area of a unit sphere in $d=6$. Similarly
\be
\Gamma^{(3)}_0(p,p')=w_0-w_0(6!g_0)^2\int^\Lambda\frac{d^6q/(2\pi)^6}
{q^2(q-p)^2(q-p')^2}\,,
\ee
again $\propto\log\Lambda$. Multiplying by $Z_\psi^{-3/2}$ the gives $w_R$. We then have to invert  this to find $g_0$ as a function of $w_R$ and $\Lambda$. Finally we find
\be
\beta(g_0)=-(\epsilon/2)g_0+56K_6g_0^3+\cdots\,.
\ee
The important feature of this is the appearance of a zero, corresponding to an RG fixed point, at $g_0^2={g_0^*}^2=O(\epsilon)$. We see that the RG flows go towards this fixed point for $\epsilon=6-d>0$. If we had taken the coupling constant to be real, this fixed point would not have been accessible. At the fixed point,
$[\Lambda(\p/\p\Lambda)-\gamma^*]G^{(2)}_0(q)=0$. To get physics out of this 
we need simple dimensional analysis, that $G^{(2)}_0(q)=q^{-2}$ times a function of $|q|/\Lambda$. Finally we find that, at the fixed point, $G^{(2)}_0(q)\propto |q|^{-2+\eta}$, where $\eta=-\epsilon/9+O(\epsilon^2)$. That $\eta$ can be negative mayhappen in a non-unitary theory. The $\epsilon$-expansion for the Yang-Lee universality class has been taken to five loops \cite{gracey}, and, when suitably resummed, extrapolates all the way down to $d=2$ where it agrees well with exactly known results, as well as more recent data in $d=3$ from the conformal bootstrap.\cite{hikami}.

\subsection{Discussion.}
The success of RG theory from the point of view of renormalized perturbative Euclidean quantum field theory led some theoreticians to assert that this is its natural home. Anyone who believes this to be the case is invited to read Fisher's masterly deconstruction of the idea in the published version of a talk he gave in 1996 to a group of quantum field theorists and philosophers of physics \cite{mef96}. Among other things, he cites many of the successes of RG theory that are not simply encapsulated in renormalized field theory, in particular the whole explanation of universality, which depends on the idea that RG flows actually take place in a large dimensional space of possible coupling constants, not just one or two; of crossover theory between nearby fixed points, a subject to which Fisher made many important contributions; other RG methods, such as Monte Carlo, functional RG, and so on, which often succeed where perturbative methods fail.

\section{The Universal Repulsive Gas and Yang-Lee Singularities}\label{securg}
We now turn to the problem of the repulsive gas singularity, and its relation to the Yang-Lee edge singularity, which Fisher and collaborators investigated in depth in two papers\cite{meflai, mefpark}. 

Consider a classical gas with purely repulsive two-body interactions $V(r)\geq0$. In that case the Mayer function $e^{-V}-1$ is negative, and it follows that the expansion of the logarithm of the grand partition function (the pressure times the volume), in powers of the activity $z$, alternates in sign with real coefficients. Since it is known to converge for small enough $|z|$, it follows that if it has a finite radius of convergence this is given by a singularity on the negative real $z$-axis. 

Moreover, this singularity appears to be universal, in the sense of the RG: that is while its location is model-dependent, the critical exponents and other critical amplitude ratios are not. This was first proposed by 
Poland\cite{poland} on the basis of cluster expansions of lattice and continuum fluids, and further confirmed by Baram and Luban\cite{baram}.
The relevant exponent is defined by the singular part of the pressure $p(z)_{\rm sing}\propto (z-z_c)^\phi$, where $\phi$ was estimated to take the values $\frac12,\frac56,\approx 1.06,\approx 1.2,\approx 1.3$ for $d=1,2,3,4,5$ respectively. The first comes from the simple calculation outlined in Sec.~\ref{sec1d}, and the second, highly nontrivial result from Baxter's solution of the hard hexagon model\cite{baxter}, which agrees with extensive recent series analysis by Jensen\cite{jensen} of hard squares. 

Fisher and Lai\cite{meflai} suggested that the origin of this universality lies in its equivalence to another problem which also exhibits a wide degree of universality: the problem of the Yang-Lee edge which Fisher had studied 17 years previously.\cite{mefyl} On this basis they conjectured that  the exponent $\phi(d)$ of the universal repulsive singularity is related to $\sigma(d)$ giving the behavior of the density of Yang-Lee zeroes by $\phi(d)=\sigma(d)+1$. The agreement, backed up by further numerical analysis of different models, was impressive (see Table 1 of their paper for an example of Fisher's thoroughness.) 

However in this paper they were not able to devise an  analytic argument for the correspondence between these problems. This gap was quickly filled in a paper with Park\cite{mefpark}, where they showed that both of these critical points appear in the same theory and are described by a $\psi^3$ field theory with purely imaginary coupling. Their arguments make for rather heavy algebra, however, and below we give a more straightforward derivation.  

\subsection{Unified field theory of the critical points.}
In this section we give a unified treatment of the Ising, Yang-Lee and repulsive critical points. We start from the Hubbard-Stratonovich representation for the partition function of the Ising model in a uniform magnetic field
\be
Z=\int[d\phi]e^{\int[-\frac12\phi\circ V^{-1}\circ\phi+\log\cosh(\phi+h)]d^dr}\,,
 \ee
where $V^{-1}$ is the positive definite matrix inverse of the interaction, and $\circ$ denotes a convolution.

The first step in a perturbative RG analysis is to identify the critical points of the Landua-Ginzberg-Wilson action above. The first variation gives
\be
V^{-1}\circ\phi=\tanh(\phi+h)\,,
\ee
which of course is the mean field equation, identifying $\phi$ as the internal magnetic field. Assuming translational invariance,
\be
\phi=V_{q=0}\cdot\tanh(\phi+h)\,.
\ee
At a critical point the hessian of second order variations vanishes. In $q$-space this is simple since it is diagonal, and assuming only ferromagnetic ordering, only the $q=0$ components are important, and we get
\be
1=V_0\,{\rm sech}^2(\phi+h)=V_0(1-V_0^{-2}\phi^2)\,.
\ee
Solving for $\phi$, we find after a little algebra
\be
V_0=\cosh^2u\,,\quad \phi=(1/2)\sinh(2u)\,,\quad h=u-(1/2)\sinh(2u)\,,
\ee
where $u=\phi+h$. 

This gives a regular parametrization (uniformization) of the mean field critical surface in terms of the complex parameter $u$ which lives on the cylinder $|{\rm Im}\,u|\leq\pi/2$. The physical cases of interest arise where this manifold intersects the real $V_0$ axis, that is
\begin{itemize}
\item $u$ purely imaginary, so $0<V_0<1$, $h$ pure imaginary: Ising model above critical temperature in purely imaginary magnetic field = Yang-Lee edge
\item ${\rm Im}\,u=\pm i\pi/2$, $V_0<0$, $z=e^{2h}<0$ with two solutions, one with $|z|<1$ = repulsive gas singularity
\item $u$ real, $V_0>1$, $h\not=0$ = the  spinodal line in a metastable low temperature phase 
\item $u=0$: the Ising critical point(in this context a multicritical point)
\end{itemize}
In all these cases the cubic term in the fluctuations has a coefficient $2V_0^{-2}\phi$. However as we move around the manifold we need to rotate the contour in $\phi$ so that the $V_0$ term remains positive, thus $\phi\to V_0^{-1/2}\phi=\sinh u$. 
This is purely imaginary at the Yang-Lee and repulsive singularities, thus confirming their identification near their critical points. Elsewhere on the critical manifold we need to choose the correct phase for $h$ so as to generate a purely imaginary cubic coupling and flow to the Fisher fixed point. It vanishes as expected at the Ising critical point.

\section{Branched Polymers and Dimensional Reduction}\label{secbp}
One of the most remarkable features of the Yang-Lee/Universal Repulsive singularity is that it also describes critical phenomena in higher dimensions, related to the counting of self-avoiding geometrical objects on the lattice or in the continuum. Specifically these have the form of trees or, more physically, branched polymers, although, if loops are included, they may be argued to be irrelevant in the RG sense. That the counting of such objects in $d$ dimensions should be related to the Yang-Lee singularity in $d-2$ dimensions was first suggested by Parisi and Sourlas\cite{ps2} on the basis of field theoretic formulations of each problem. They observed an emergent supersymmetry in the leading diagrams whereby two anticommuting coordinates canceled two of the usual commuting euclidean coordinates and led to the aforesaid correspondence. There is ample numerical evidence for this, see for example Table 1 of Fisher and Lai\cite{meflai}. However it is based on perturbative RG ideas and as such could be questioned, especially when a similar earlier argument by Parisi and Sourlas\cite{ps1} for the random field Ising model fails in low enough dimensions\cite{imbrie}.\footnote{Various explanations have been offered for this failure. However it may be accounted for by observing that dimensional reduction applies only to the tree diagrams, which describe the long-range limit of the random field Ising model.\cite{jcinprep}}

However  in  2001-02 Brydges and Imbrie\cite{bi1,bi2}  devised a class of models for which the connection between the repulsive singularity and branched polymers in two higher dimensions, which is both rigorous and appealing in its physical description. Although their arguments ultimately appeal to supersymmetry, we shall give a simpler version based on Kirchhoff's matrix-tree theorem of 1847.\cite{kirchhoff} 

Consider a simple classical gas in $d$ dimensions with repulsive two-body interactions. The grand partition function is
\be
\Xi(z)=\sum_{N=0}^\infty\frac{z^N}{N!}\prod_{i=1}^N\int d^dr_i\prod_{1\leq j<k\leq N}e^{-V(r_j-r_k)}\sim e^{p(z)|\Omega|}\,,
\ee
where $z$ is the activity, $p$ is the pressure  and $|\Omega|$ is the total volume, assumed large. 

The cluster expansion proceeds by writing $e^{-V}=1+f$ and expanding in powers of the Mayer function $f$. This gives a sum over subgraphs which span the complete graph with $N$ vertices, each edge of the subgraph carrying a factor $f(r_j-r_k)$. Taking the logarithm of this expansion gives a sum over connected diagrams, and the mean density $\langle n\rangle=z(d/dz)
\log\Xi$ is given by a sum over \em rooted \em connected spanning subgraphs $C$ of the complete graph $K_N$:
\be
\langle n\rangle_d=\sum_{N=1}^\infty\frac{z^N}{(N-1)!}\sum_C
\prod_{i=}^N\int d^dr_i\delta^d(r_1)\prod_{jk\in e(C)}f(r_j-r_k)\,,
\ee
where $e(C)$ denotes the set of edges of $C$. 

Now suppose for simplicity that $f(r_j-r_k)=e^{-\alpha(r_j-r_k)^2}$ (a more general dependence may be written as a superposition of such terms.) Then
\be
\prod_{jk\in e(C)}f(r_j-r_k)=e^{-\sum_{\mu=1}^dr^\mu_jA_{jk}r^\mu_k}\,,
\ee
where $A_{jk}=A_{kj}=-\alpha$ if $(kj)\in e(C)$, and $A_{jj}=-\sum_{k\not=j}A_{jk}$, that is $A$  is the laplacian on $C$. For a fixed $C$, the gaussian integral over $r_2,\ldots, r_N$ gives 
\be\label{fact}
\pi^{(N-1)d/2}(\det A')^{-d/2}=\pi^{-(N-1)}(\det A')\times 
\pi^{(N-1)(d+2)/2}(\det A')^{-(d+2)/2}\,,
\ee
where $A'$ is the matrix formed from $A$ by striking out the first row and column. 
According to the matrix-tree theorem\cite{kirchhoff} $\det A'$ can be written as a sum over rooted spanning trees $T$
\be
\det A'=\sum_{T\subseteq C}\prod_{jk\in T}A_{jk}\,,
\ee
where in this case $A_{jk}=-\alpha$.

The second factor in (\ref{fact}) is the weight of the cluster diagram in $d+2$ dimensions.
This may be written back in terms of $e^{-V(r_j-r_k)}=1+e^{-\alpha (r_j-r_k)^2}$.
The  outcome is that
\be
\langle n(z)\rangle_d=
z\sum_{N=1}^\infty\frac{(-\alpha z/\pi)^{N-1}}{(N-1)!}C_N\prod_i\int d^{d+2}r_i\delta^{d+2}(r_1)\prod_{1\leq j<k\leq N}(1+e^{-\alpha (r_j-r_k)^2})\,,
\ee
where $C_N=N^{N-1}$ is the number of rooted spanning trees on $K_N$.
This may be viewed as the partition function, or generating function, for branched polymers in $d+2$ dimensions, with a repulsive weight for each pair of monomers, or vertices of the graph. More generally we have\cite{bi1}
\be\label{nz}
\langle n(z)\rangle_d=
z\sum_{N=1}^\infty\frac{(-z/\pi)^{N-1}}{(N-1)!}\prod_i\int d^{d+2}r_i\delta^{d+2}(r_1)\sum_T\prod_{jk\notin e(T)}P(r_{jk}^2)\prod_{jk\in e(T)}Q(r_{jk}^2)\,,
\ee
where $P(r^2)=e^{-V(r)}$ and $Q(r^2)=P'(r^2)$. For a reasonable potential $V$, $P<1$ and $P\to 1$ as $r\to\infty$, corresponding to a short-range repulsive weight for pairs of  monomers which are not neighbors on the tree, and $Q>0$ with $Q\to0$ at infinity, giving a weight which is peaked at some value of $r^2$, corresponding to finite-distance attraction and short-distance steric repulsion between neighboring monomers. 

(\ref{nz}) may be integrated with respect to $z$ to give
\be
p(z)_d=-\pi\sum_{N=1}^\infty(-z/\pi)^N{\cal Z}(N)_{d+2}\,,
\ee
where now ${\cal Z}(N)_{d+2}$ is the partition function for unrooted branched polymers with $N$ monomers. This is the main result of Brydges and Imbrie.\cite{bi1}
It may be extended to correlation functions.\cite{bi2} The change of sign of the activity is important, because it implies that the repulsive gas singularity at $z=-z_c$ controls the large $N$ behavior of ${\cal Z}(N)_{d+2}$. Explicitly, if $p(z)_d$ has a singular piece $\propto (z+z_c)^{\phi(d)}$, then
\be
{\cal Z}(N)_{d+2}\sim N^{-1-\phi(d)}(\pi z_c)^{-N}\,.
\ee

\subsection{Directed branched polymers}

Similar reasoning may be applied to the problem of directed branched polymers or trees, in which each edge is preferentially oriented in a particular spatial direction, labeled by a coordinate, say $t$. In this case the correspondence is between the branched polymer problem and the relaxational dynamics of the repulsive gas. This was first conjectured using a field theory formulation\cite{jcdbp}, and Dhar\cite{dhardbp} showed an exact correspondence between a discrete  dynamical model and hard squares and hexagons. Subsequently Imbrie\cite{imdbp}, using methods similar to that in Ref.~\onlinecite{bi1}, constructed a model for which the correspondence is exact.
Here we present a more refined version of Ref.~\onlinecite{jcdbp}.

Let us begin with the sine-Gordon transformed version of the repulsive lattice gas partition function 
\be
\Xi=\int[d\phi]e^{-{\cal H}}=\int[d\phi]e^{-\frac12\sum_{r,r'}\phi(r)V^{-1}(r,r')\phi(r')+z\sum_re^{i\phi(r)}}\,,
\ee
where $V^{-1}$ is the matrix inverse of the 2-body potential. The derivation of this is similar to that of (\ref{hs}), except that we sum over occupation numbers $n(r)=0,1,2,\ldots$, and the factor $i$ necessary to make the gaussian integral converge.

Now we take $\cal H$ as an effective Landau-Ginzberg-Wilson free energy, and consider its
relaxational 'model A' dynamics,\footnote{Although this comes from a fluid whose density is conserved and which therefore should satisfy model B dynamics\cite{tauber},
the field $\phi$ is essentially the local chemical potential and is not conserved.}
defined by the stochastic equation\footnote{This is the physicist's crude way of writing $d\phi_t(r)=\ldots+D^{1/2}dB_t(r)$ where $B_t(r)$ are independent Brownian motions.}
\be\label{phidot}
\frac{\partial\phi(r)}{\partial t}=-\Gamma\frac{\partial{\cal H}}{\partial\phi(r)}
+\eta(r,t)\,,
\ee
where $\Gamma$ is a frictional coefficient and$\eta$ is a brownian noise with covariance $\langle\eta(r,t)\eta(r',t')\rangle=2D\delta_{r,r'}\delta(t-t')$. From the corresponding Fokker-Planck equation one may show that as long as $D=\Gamma kT$ the joint probability distribution of $\{\phi(r)\}$ relaxes to a thermal one $\propto e^{-{\cal H}/kT}$.

To proceed we  use the response function formalism\cite{tauber}, writing a delta function of (\ref{phidot}) as
\be
\int[d\bar\phi]e^{\int dt\sum_r\bar\phi(r,t)[\partial_t\phi(r,t)+\Gamma\partial{\cal H}/\partial\phi(r)-\eta(r,t)]}\,.
\ee
In this form it is straightforward to average over the noise to get a term $D\bar\phi^2$ in the action.
Physical observables are then defined by adding source terms.

In our case the response functional is
\be\label{Sresp}
S=\int dt\sum_r\big(\bar\phi(r,t)[\partial_t\phi(r,t)+\Gamma\sum_{r'}V^{-1}(r,r')\phi(r')-ize^{i\phi(r,t)}]-D\bar\phi(r,t)^2\big)\,.
\ee

Let us now compare with a field theory derived from an explicit lattice model of directed branched polymers. We use a variant of the sine-Gordon transformation first deployed in Ref.~\onlinecite{jcdbp}. On each site $(r,t)$ of a directed lattice introduce a commuting ring consisting of complex linear combinations of pseudospins  $(a(r,t),\bar a(r,t))$, satisfying the algebra 
\be
a^2=a\,,\qquad \bar a^2=0\,,
\ee
together with a linear mapping Tr to the complex numbers defined by
\be
{\rm Tr}\,1=1\,,\qquad{\rm Tr}\,a=0\,,\qquad {\rm Tr}\,\bar a=1\,,\qquad {\rm Tr}\,\bar aa=1\,.
\ee
Consider the expression
\be
{\rm Tr}\,\bar a(0,0)\prod_{r,r'}\prod_{t'>t}\big(1+xV(r,t;r',t')\bar a(r',t')a(r,t)\big)\,.
\ee
Expanding this out in powers of $x$, each term is associated with a subgraph of the lattice, and using the above rules the operation Tr projects onto directed trees rooted at $(0,0)$, so gives the generating function for these. Since $\bar a^2=0$ this can be exponentiated and written as a gaussian integral
\be
\exp\big(x\sum_{r,r',t,t'}V(r,t;r',t')\bar a(r',t')a(r,t)\big)
=\int[d\bar\phi][d\phi]e^{\sum_{r,t}x^{-1}\bar\phi V^{-1}\phi
+i\bar a\bar\phi+ia\phi}\,.
\ee
The operation Tr may now be performed to give
a functional
\be
S'=\int dt\sum_r\big(\bar\phi(r,t)[\partial_t\phi(r,t)+\Gamma\sum_{r'}V^{-1}(r,r')\phi(r')+\log(1+ix\bar\phi\, e^{i\phi})\big)\,,
\ee
where we have rescaled $\bar\phi$ and assumed that $V$ is short-ranged in $t$ so we may make a derivative expansion in $\partial_t$. Now expanding the logarithm and keeping relevant terms we get (\ref{Sresp}) with negative activity $z=-x$.

Some of the critical exponents of directed branched polymers may now be inferred from the theory of dynamic critical behavior\cite{tauber}. The Fourier transform $G(\omega,q)$ of the response function $\langle\phi(r,t)\bar\phi(0,0)\rangle$. which gives the local density of the branched polymer at $(r,t)$, when evaluated at $\omega=0$, should equal the static correlation function of the repulsive singularity. Thus $G(0,0)\sim (x-x_c)^{-\gamma}$ and $\xi_\perp\sim (x-x_c)^{-\nu}$ where $\xi_\perp$ is the correlation length in the $r$ directions, and $\gamma,\nu$ are the exponents of the repulsive gas singularity, or, equivalently, the Yang-Lee singularity, in $d$  dimensions. By our earlier arguments in Sec.~\ref{secmomrg}, $\nu=1/y_h$ and $\gamma=(2y_h-d)/y_h$. 
By hyperscaling the singularity in the free energy is $(x-x_c)^{-1+\sigma(d)}$, and therefore the number of directed branched polymers of mass $N$ in $d+1$ dimensions grows like
$N^{1+\sigma(d)}x_c^{-N}$. Note that these arguments give no immediate information about the correlation length $\xi_t$, as this requires knowledge of the dynamic exponent of the Yang-Lee problem.

\section{Exact Results in Two dimensions}\label{exact2d}
The scaling limit of the Yang-Lee/universal repulsive gas singularity at or near the critical point has been a fertile paradigm for theoretical investigations of two dimensional critical behavior because of (i)  its simplicity: it has only one relevant scaling field, making it simpler in some ways than the Ising model, although it is not a free theory in any sector, as is the  Ising model; (ii) it gives the simplest example of a non-unitary theory, albeit not exhibiting all the features of which more general such theories are capable. Unfortunately we have space for only a brief description, and refer the reader to the literature for more details.
\subsection{Conformal field theory}
At a rotationally invariant critical point with short range interactions, scale invariance is usually enlarged to conformal symmetry, which may loosely be thought of as invariance (or covariance,  as Fisher always insisted upon saying) under non-uniform scaling. In two dimensions it is very powerful, as any complex analytic function generates a locally conformal mapping of the complex plane, and even in higher dimensions. once the kinematic clutter was tidied away, the conformal bootstrap (described elsewhere in this volume \cite{hikami}) has become the most effective, numerically and analytically. The ccnformal field theory of the Yang-Lee edge singuarity in two dimensions was identified \cite{jcylcft} early on, shortly after Belavin, Polyakov and Zamolodchikov\cite{bpz} formulated their classification of the so-called \em minimal \em models, for which the number of irrelevant fields grows less fast than generically, owing to an abundance of redundant fields, or null fields, which might be expected to occur in polynomial field theories like $i\psi^3$.  

The minimal models are characterized by two co-prime positive integers $(p,p')$. The scaling fields  are organized into representations of the Virasoro algebra, labelled by integers $(r,s)$, within which the scaling dimensions are spaced by integers. The most relevant field $\phi_{r,s}$ in each representation has scaling dimension
\be\label{xrs}
x_{r,s}=\frac{(rp-sp')^2-(p-p')^2}{2pp'}=x_{p'-r,p-s}\,,
\ee
where $1\leq r\leq p'-1$, $1\leq s\leq p-1$. This dual appearance of each field and the existence of null fields restricts the allowed structure of  the operator product expansion to be
\be
\phi_{r,s}\cdot\phi_{r',s'}=\sum_{|r-r'|\leq r''\leq r+r'-1}\,\sum_{|s-s'|\leq s''\leq s+s'-1}\phi_{r'',s''}\,.
\ee
Assuming that the Yang-Lee conformal field theory is a minimal model, it must satisfy: (a) there is only one relevant field, that is with $x_{r,s}<2$ (apart from the identity $\phi_{1,1}$); (b) the 3-point function is non-zero, that is the OPE $\phi_{r,s}\cdot\phi_{r,s}$ should contain $\phi_{r,s}$ itself, or its image $\phi_{p'-r,p-s}$. It turns out that the latter can hold with $(r,s)=(1,2)$, with
\be
\phi_{1,2}\cdot\phi_{1,2}=\phi_{1,1}+\phi_{1,3}=\phi_{1,1}+\phi_{2-1,5-2}\,,
\ee
as long as $(p,p')=(2,5)$.   Using (\ref{xrs}) this gives $x=-\frac25$ and, using Fisher's scaling relation, $\sigma=-\frac16$, in agreement with Baxter's exact result $\phi=\frac56$ for hard hexagons.\cite{baxter} 

This was the first successful application of CFT methods to a non-unitary theory. It predicts not only the leading exponent bot also the full finite-size scaling spectrum when the theory is compactified on a cylinder. Moreover, since as was argued by Belavin et al., the existence of a null field in the representation $(r,s)$ implies that the correlators satisfy linear partial differential equations of order $r\cdot s$, the 4-point function may computed exactly (using crossing, which is a main ingredient of the conformal bootstrap) and the universal coefficient of the three-point function extracted, the result being 
\be
\langle\phi(r_1)\phi(r_2)\phi(r_3)\rangle=iC^{1/2}|r_{12}r_{23}r_{31}|^{2/5}\,,
\ee
where the normalization $\langle\phi(r_1)\phi(r_2)\rangle=|r_{12}|^{4/5}$,
and
\be
C=\frac{\Gamma(\frac65)^2\Gamma(\frac15)\Gamma(\frac25)}
{\Gamma(\frac35)\Gamma(\frac45)^3}\,.
\ee
Note that the factor of $i$ in the 3-point function confirms the imaginary cubic coupling in the field theory. 
This has been verified numerically\cite{wydro}. We quote it in full to show just hoe nontrivial these types of CFT result are. 

Although the Yang-Lee CFT is non-unitary, its spectrum is real. This is a consequence of the symmetry of the associated quantum hamiltonian under simultaneous time reversal $H\to H^*$ and $\phi\to-\phi$. 

\subsection{The simplest $S$-matrix.}
Just as the universal properties at a critical point are encoded in a conformal field theory, those of the neighboring scaling region are generally encoded in a  massive field theory. To be precise, this scaling limit involves taking the short-distance cut-off, e.g. the lattice spacing, $a\to0$, while keeping the correlation length $\xi$, as measured in laboratory units, fixed. Since in general $\xi\sim ag^{-1/y}$, where $g$ is a relevant coupling with RG eigenvalue $y>0$, this limit involves letting $g\to0\propto a^y$. The mass scale of the limiting field theory is then given by the inverse correlation length $\xi^{-1}$. Note that making an RG transformation $a\to ba$ simply rescales the constant of proportionality above. Thus all points on the relevant outflow trajectory from the fixed point correspond to the \em same \em massive continuum theory, when expressed in terms of the physical masses.\footnote{Fisher was very fond of pointing out that what is often termed the so called RG in renormalized field theory, which ignores the infinitude of other couplings, is a tautology.}

A massive field theory is characterized by its particle spectrum and $S$-matrix. In general the spectrum may contain particles of different masses and quantum numbers, but, since the Yang-Lee theory has only one type of particle, we shall assume this. A single particle of mass $m$ with momentum $p^1$ has energy $p^0=\sqrt{m^2+{p^1}^2}$. It is convenient to parametrize $(p^0,p^1)=(m\cosh\theta,m\sinh\theta)$ in terms of the rapidity $\theta$. A convenient basis is then that of the asymptotic states 
$|\theta_1,\theta_2,\cdots\rangle$ which may be thought of as a multiparticle state  before the interaction is switched on. This is turned on for some long but finite time, after which the system will be in some linear superposition of such states which, because of the quantum evolution being unitary, may be written $S^{-1}|\theta_1,\theta_2,\cdots\rangle$
where $S$ is a unitary operator. As usual, scattering processes  conserve total energy and momentum, which are integrals of components of the conserved energy-momentum tensor $T_{\mu\nu}$ over space. In a CFT, however, there are extra conservation laws, which are more easily understood in light-cone coordinates $x^\pm=x^0\pm x^1$, where the conservation laws read
 \be
 \p_-T_{++}+\p_+T_{+-}=0\,,\quad \p_+T_{--}+\p_-T_{-+}=0\,.
\ee
In a 2d CFT, $T_{+-}=T_{-+}=0$, so $T_{++}$ and $T_{--}$ are separately conserved. Moreover, so are any powers $T_{++}^p$ and $T_{--}^{p}$.

Although this is not generic, it may happen that when a relevant coupling $g$ is switched on, at least some of these currents remain conserved. Zamolodchikov\cite{zamcons} argued that, for certain deformations of minimal CFTs, this is the case. The reasoning is simply that if one asks what can lie on the right hand side of the equation $\p_-T_{++}^p=?$, the only possibilities are derivatives $\p_+$ of some other local field. His argument applies to the deformation of the Yang-Lee CFT by its (only) relevant field.\cite{jcmuss} 

The consequences for the scattering are profound. It implies that not only are
$\sum_je^{\pm\theta_j}$ the same in the initial and final state, but also
$\sum_je^{\pm p\theta_j}$ for an infinite set of values of $p$. This can only happen if the $\{\theta_j'\}$ in the final state are some permutation of those in the initial state. Moreover it implies factorization of the $S$ matrix: the $n$ particle $\to n$ particle $S$ matrix can be written as a product of $2\to2$ matrices. Thus we need focus on only $S_{2\to2}(\theta_1,\theta_2)$, which may be written $S(\theta_1-\theta_2)$ by Lorentz invariance. 

The usual constraints of analyticity, unitarity and crossing are simple in the rapidity basis: 
\begin{itemize}
\item $S(\theta)$ is analytic in $0<{\rm Im}\,\theta<\pi$, apart from possible poles on the imaginary axis corresponding to bound states;
\item unitarity $S(\theta)S(-\theta)=1$;
\item crossing symmetry $S(\theta)=S(i\pi-\theta)$.
\end{itemize}
Moreover, we want three 3-point coupling to be non-zero, that is the particle should occur as a bound state in the 2-particle spectrum. By considering $3\to3$ scattering we may go to the pole in, say the (12) channel and find its residue. this gives a non-linear relation for the $2\to2$ $S$ matrix, called the \em bootstrap \em condition\footnote{This is the original meaning of the term, not to be confused with the conformal bootstrap.}
$S(\theta)=S(\theta-\frac13i\pi)S(\theta+\frac13i\pi)$.

The minimal solution to these conditions is then\cite{jcmuss}
\be
S(\theta)=\frac{(e^\theta-e^{-2\pi i/3})(e^\theta-e^{-i\pi/3})}
{(e^\theta-e^{2\pi i/3})(e^\theta-e^{i\pi/3})}\,.
\ee

This simple expression encodes all the properties of the scaling limit away from the critical point. For example, it may used to construct the form factors of local operators $\langle0|\Phi|\theta_1,\theta_2,\ldots\rangle$, which may then be sewn together using unitarity to find highly convergent  expressions for the two-point functions. It may also be used as input to the thermodynamic  Bethe ansatz to extract the CFT data, and check the consistency of the whole circular argument.

\section{Conclusions}
The Yang-Lee edge problem, in addition to being an important way of understanding criticality in Ising and other systems, is in itself a wonderful playground to explore criticality in general, and in particular systems with complex weights, or equivalently non-unitary field theories, which arise in many areas of condensed matter physics. The connections with  polymer problems in higher dimensions are a bonus. It is not surprising that Michael Fisher found it  fascinating, returning to it more than once.

This work was prepared while the author was attached to the University of California, Berkeley.

.


\begin{thebibliography}{99}   
%
\bibitem{YL} C.~N. Yang and T.~D. Lee, Statistical Theory of Equations of State and Phase Transitions I. Theory of Condensation, \emph{Phys. Rev.} {\bf 87} (3), 404-409 (1952).
%
\bibitem{LY} T.~D. Lee and C.~N. Yang, Statistical Theory of Equations of State and Phase Transitions II. Lattice Gas and Ising Model, \emph{Phys. Rev.} {\bf 87} (3), 410-419 (1952).
%
\bibitem{onsager} L. Onsager, Crystal Statistics. I. A Two-Dimensional Model with an Order-Disorder Transition,
\emph{Phys. Rev.} {\bf 65}, 117-149 (1944).
%
\bibitem{yangmag} C.~N. Yang, The Spontaneous Magnetization of a Two-Dimensional Ising Model, \emph{Phys. Rev.} {\bf 85}, 808-816 (1952).
%
\bibitem{xz} H.-L. Xu and A.~B. Zamolodchikov, 2D Ising Field Theory in a magnetic field: the Yang-Lee singularity, \emph{Journal of High Energy Physics} {\bf 08} 57 (2022).
  %
\bibitem{kg} P.~J. Kortman and R.~B. Griffiths, Density of Zeros on the Lee-Yang Circle for Two Ising Ferroinagnets. \emph{Phys. Rev. Lett.} {\bf 27} (21) 1439-1442 (1971).
%
\bibitem{mefyl} M.~E. Fisher, Yang-Lee Edge Singularity and $\phi^3$ Field Theory, \emph{Phys. Rev. Lett.}  {\bf 40} (25) 1610-1613 (1978).
%
\bibitem{rue} D.~Ruelle, Characterization of Lee-Yang polynomials, \emph{Ann. of Math.} {\bf 171} 589?603 (2010).
%
\bibitem{mefyl1d} M.~E. Fisher, Yang-Lee Edge Behavior in One-Dimensional Systems, \emph{Prog. Theoretical Physics Supp.}, {\bf 69}, 14-29 (1980).
%
\bibitem{wf} K.~G. Wilson and M.~E. Fisher, Critical Exponents in 3.99 Dimensions, \emph{Phys. Rev. Lett.}  {\bf 28} (4) 240-243 (1972).
%
\bibitem{wilson1} K.~G. Wilson, Renormalization group and critical phenomena. I. Renormalization group and the Kadanoff scaling picture, \emph{Phys. Rev. B} {\bf 4} (9), 3174-3183 (1971).
%
\bibitem{wilson2} K.~G. Wilson, Renormalization group and critical phenomena. II. Phase-Space Cell Analysis of Critical Behavior, \emph{Phys. Rev. B} {\bf 4} (9), 3184-3205 (1971).
%
\bibitem{jcbook} J. Cardy, {\it Scaling and Renormalization in Statistical Physics}.
(Cambridge University Press, Cambridge, 1996) pp. 45-55.
%
\bibitem{gracey} M. Borinsky, J.~A. Gracey, M.~V. Kompaniets and O. Schnetz,
Five loop renormalization of $\phi^3$ theory with applications to the Lee-Yang edge singularity and percolation theory, \emph{Phys. Rev. D} {\bf 103}, 116024 (2021).
%
\bibitem{hikami} S. Hikami, Conformal Bootstrap Analysis for Yang-Lee Edge Singularity, \emph{Prog. Theoretical and Experimental Physics}, {\bf 2018} (5),  053101 (2018).
%
\bibitem{mef96} M.~E. Fisher, Renormalization group theory: Its basis and formulation in statistical physics, \emph{Rev. Mod. Phys.} {\bf 70} (2), 653-681 (1998).
%
\bibitem{meflai} S.-N. Lai and M.~E. Fisher, The universal repulsive-core singularity and Yang?Lee edge criticality, \emph{J. Chem. Phys.} {\bf 103}, 8144-8155 (1995). 
%
\bibitem{mefpark} Y. Park and M.~E. Fisher, 
Identity of the universal repulsive-core singularity with Yang-Lee edge criticality,
\emph{Phys. Rev. E} {\bf 60} (6), 6323-6328 (1999).
%
\bibitem{poland}  D. Poland, On the universality of the nonphase transition singularity in hard-particle systems, \emph{J. Stat. Phys.} {\bf 35}, 341-353 (1984).
%
\bibitem{baram} A. Baram and M. Luban, Universality of the cluster integrals of repulsive systems, \emph{Phys. Rev. A} {\bf 36}, 760-765 (1987).
%
\bibitem{baxter} R.~J. Baxter, Hard hexagons: exact solution,
\emph{J. Phys. A: Math. Gen.} {\bf 13}, L61 (1980).
%
\bibitem{jensen} I. Jensen, Comment on ?Series expansions from the corner
transfer matrix renormalization group method: the
hard-squares model?, \emph{J. Phys. A: Math. Gen.} {\bf 45}, 508001 (5pp.) (2012).
%
\bibitem{ps2} G. Parisi and N. Sourlas, Critical Behavior of Branched Polymers and the Lee-Yang Edge Singularity, \emph{Phys. Rev. Lett.} {\bf 46}, 871-874 (1981).
%
\bibitem{ps1} G. Parisi and N. Sourlas, Random Magnetic Fields, Supersymmetry, and Negative Dimensions, \emph{Phys. Rev. Lett.} {\bf 43}, 744-745 (1979).
%
\bibitem{imbrie} J.~Z. Imbrie, Lower critical dimension of the random-field Ising model, \emph{Phys.Rev. Lett.} {\bf 53}, 1747?1750, 1984.
%
\bibitem{jcinprep} J. Cardy, The liquid-gas critical point in a random medium and dimensional reduction, in preparation.
%
\bibitem{bi1} D.~C. Brydges and J.~Z. Imbrie, Branched Polymers and Dimensional Reduction, \emph{Annals of Mathematics} {\bf 158}, 1019-1039 (2003).
%
\bibitem{bi2} D.~C. Brydges and J.~Z. Imbrie, Dimensional Reduction Formulas for Branched Polymer Correlation Functions, \emph{J. Stat. Phys.} {\bf 110}, 503-518 (2003).
%
\bibitem{kirchhoff} G. Kirchhoff, Ueber die Auflijsung der Gleichungen, auf welche man bei der Untersuchung der linearen Vertheilung Galvanischer Str\"ome gef\"uhrt wird, \emph{Ann. Phys. Chem.} {\bf 72}, 497-508 (1847).
%
\bibitem{jcdbp}  J. Cardy, Directed lattice animals and the Lee-Yang edge singularity, \emph{J. Phys. A: Math. Gen.} {\bf 15} L593-595 (1982).
%
\bibitem{dhardbp} D. Dhar, Exact Solution of a Directed-Site Animals-Enumeration Problem in Three Dimensions.
\emph{Phys. Rev. Lett.} {\bf 51}, 853-856 (1983); Erratum \emph{Phys. Rev. Lett.} {\bf 51}, 1499 (1983).
%
\bibitem{imdbp} J.~Z. Imbrie, Dimensional Reduction for Directed Branched Polymers, \emph{J. Phys. A: Math. Gen.} {\bf 37}, L137--L142 (2004).
%
\bibitem{tauber} U.~C. T\"auber, Critical Dynamics (Cambridge University Press, Cambridge \& New York, 2014).
%
\bibitem{jcylcft} J. Cardy, Conformal Invariance and the Yang-Lee Edge Singularity in Two Dimensions, \emph{Phys. Rev. Lett.} {\bf 54} (13), 1354-1356 (1985). 
%
\bibitem{bpz} A.~A. Belavin, A.~M. Polyakov and A.~B. Zamolodchikov, Infinite conformal symmetry of critical fluctuations in two dimensions, \emph{J. Stat. Phys.} {\bf 34}, 763 (1984).
%
\bibitem{wydro} T. Wydro and J.~F. McCabe,
Tests of conformal field theory at the Yang?Lee singularity,
\emph{AIP Conference Proceedings} {\bf 1198}, 216 (2009).
%
\bibitem{zamcons} A.~B. Zamolodchikov, lntegrable field theory from conformal field theory, (Proc. Taniguchi Symp. (Kyoto, 1988)) \emph{ Advanced Studies in Pure Mathematics} {\bf 19}, 641-674, (1989). 
%
\bibitem{jcmuss} J. Cardy and G. Mussardo, S-matrix of the Yang-Lee edge singularity in two dimensions, \emph{Phys. Lett. B}, {\bf 225} (3), (1989). 

   

\end{thebibliography}
\end{document}